\documentclass[11pt]{article}
\usepackage{epsfig}
\usepackage{graphicx}
\usepackage{amsmath}
\usepackage{amssymb}
\usepackage{amsxtra}
\usepackage{amsfonts,amsthm}
\def\Journal#1#2#3#4{{#1} {\bf #2} (#4) #3 }
\def\IJMPA{{ Int. J. Mod. Phys.}  A}

\def\BWP{ Bled Workshops in Physics}

\def\({\left(}
\def\){\right)}

\def\beq{\begin{equation}}
\def\eeq{\end{equation}}
\def\lsim{\mathrel{\rlap{\lower4pt\hbox{\hskip1pt$\sim$}}
    \raise1pt\hbox{$<$}}}         %less than or approx. symbol

\def\gsim{\mathrel{\rlap{\lower4pt\hbox{\hskip1pt$\sim$}}
    \raise1pt\hbox{$>$}}}         %greater than or approx. symbol
\title{10 years of dark atoms of composite dark matter}
\author{
M.Yu.~Khlopov $^{a,b}$\footnote{khlopov@apc.univ-paris7.fr}
}
\begin{document}
%%%%%%%%%%%%%%%%%%%%%%%%%%%%Title page
\maketitle
\begin{flushleft}{{\it $^{a}$ \small National Research Nuclear University "MEPHI" (Moscow Engineering Physics
Institute), 115409 Moscow, Russia \\
Centre for Cosmoparticle Physics ``Cosmion'' 115409 Moscow, Russia
\\ $^{b}$ \small APC laboratory 10, rue Alice Domon et Lonie Duquet
75205 Paris Cedex 13, France}}
\end{flushleft}
\begin{abstract}
In 2005 Sheldon Glashow has proposed his sinister model, opening the path to composite-dark-matter scenarios, in which heavy stable electrically
charged particles bound in neutral atoms play the role of dark matter candidates. Though the general problem of new stable single charged particles, forming with ordinary electrons anomalous isotopes of hydrogen, turned out to be unresolvable in Glashow's scenario, this scenario stimulated development of composite dark matter models, which can avoid the trouble of anomalous isotope overproduction.
In the simplest case composite dark matter may consist of -2 charged particles, bound by ordinary Coulomb interaction with primordial helium in OHe dark matter model.
The advantage and open problems of this model are discussed.
\end{abstract}

\section{Introduction}
The existence of dark matter, constituting dominant fraction of the matter content of the Universe,
is one of the cornerstones of the modern cosmology, but its physical nature is still elusive.
The results of direct searches for dark matter are reviewed in \cite{DAMAtalk}. Though the apparent
contradiction of these results comes from the uncritical comparison of the data, obtained with the use of different techniques, and even their interpretation in the terms of Weakly Interacting Massive Particles (WIMPs) is still not ruled out \cite{DAMAtalk},
a more general approach to a possible solution of the dark matter problem is appealing. Here we concentrate on a possibility that in the same way as the ordinary matter is composed by atoms, which consist of electrically charged electrons and nuclei, bound by Coulomb forces, new electrically charged stable particles may be bound by ordinary Coulomb field in the dark atoms of the dark matter. The electrically charged constituents of dark atoms may be not only elementary particles, but can be composite objects, as are ordinary nuclei and nucleons.

In 2005 the idea of such a multi-composite dark matter was put forward by Sheldon Glashow in his sinister model \cite{Glashow}. The model assumed a set of million times heavier partners of ordinary quarks and leptons related by a strict symmetry. The lightest of these partners (tera-electron and tera-U-quark) were stable and could form a stable tera-helium atom (UUU)EE, in which +2 charged quark cluster (UUU) was bound by ordinary Coulomb force with two tera-electrons.
It was proposed that in the early Universe the excessive U-quarks first bind in (UUU) cluster, which recombines then with excessive tera-electrons to form tera-helium atom. The unrecoverable problem of this scenario, revealed in \cite{BKSR1}, was inevitable overproduction of +1 and +2 charged remnants of incomplete binding, like (Uud), (UUu) hadrons or (UUU)E ions, which bind with ordinary electrons in atomic states that look like anomalous isotopes of hydrogen and helium. Moreover all the free tera-electrons turned to bind with primordial helium, as soon as it was formed in Big Bang Nucleosynthesis, in +1 charged ion
(EHe), increasing the list of undesirable +1 charged species and preventing any possible reduction of their abundance. It makes impossible to realize the dark atom scenario not only in Glashow's sinister model, but also in any other model predicting stable +1 and -1 charged species. However these studies stimulated further development of the idea of composite dark matter particles both in the form of stable heavy quark clusters and dark atoms, in which new stable charged particles are bound.

Starting from 2006 various realizations of possible solution for dark atom scenario were proposed \cite{invention,KK1,ac,5g,spectro,DMRev,DDMRev}, in which the important role of stable -2 charged species was revealed. These species are bound with primordial helium in neutral OHe atoms, which play important catalyzing role in reduction of all the undesirable positively charged heavy species that can give rise to anomalous isotopes, as well as can be the candidate for composite dark matter, dominating in the matter density of the Universe. Such candidates for dark matter should consist of
negatively doubly-charged heavy (with the mass $\sim 1$ TeV) particles, which are called O$^{--}$, coupled
to primordial helium. Lepton-like technibaryons, technileptons, AC-leptons or clusters of
three heavy anti-U-quarks of 4th generation with strongly suppressed hadronic
interactions are examples of such O$^{--}$ particles (see \cite{invention,KK1,ac,spectro,DMRev,DDMRev} for
a review and for references). Another direction of composite dark matter scenario is to consider
neutral stable heavy quark clusters as it is proposed in the approach of \cite{Norma}. However, even in this approach heavy stable -2 charged clusters ($\bar u_5 \bar u_5 \bar u_5$) of stable antiquarks $\bar u_5$ of 5th generation can also find their physical basis \cite{5g}.

Here we briefly outline the advantages of the OHe dark atom scenario in its ability to explain some puzzles of direct and indirect dark matter searches, specifying collider and non-collider probes for this scenario as well as its open problems.
\section{OHe solutions for puzzles of dark matter searches}
It is assumed that together with generation of baryon asymmetry the excess of O$^{--}$ particles is generated. This assumption finds natural basis
in Walking Technicolor models, which provide proper ratio of baryon and O$^{--}$ excess due to sphaleron transitions in the early Universe.
Similar relationship can take place in any other model with O$^{--}$ coupling to electroweak sphalerons.

As soon as primordial helium is produced in Big Bang nucleosynthesis, it captures all the free O$^{--}$ forming OHe atoms. These atoms catalyze binding and annihilation of all the undesirable positively charged species. Before OHe gas starts to dominate on matter dominated stage, it decouples from plasma and radiation, what is necessary for its role of proper dark matter candidate. It leads to scenario of Warmer than Cold Dark Matter (WtCDM) with a slight suppression of small scale fluctuations. This suppression is
less pronounced than in the Warm Dark Matter scenario, but still it can be of interest for solution of small scale cusp problem of the standard CDM. In spite of its strong nuclear interaction OHe gas is collisionless at galactic scale,
but all the dense matter objects like stars or planets are opaque for it. Due to this opacity the infalling flux of OHe is captured and thermalized in
the terrestrial matter.

It is assumed that the effective
potential between OHe and a normal nucleus would have a barrier, preventing He and/or
O$^{--}$ from falling
into the nucleus. Under these conditions elastic collisions dominate in OHe
interactions with matter, and lead to a successful OHe scenario.
The cosmological and astrophysical effects of such composite dark matter (dark atoms of
OHe) are dominantly related to the helium shell of OHe and involve only one parameter
of new physics $-$ the mass of O$^{--}$.
\subsection{OHe solution for puzzles of direct dark matter search}
Dark atom interpretation of the puzzles of direct dark matter search is based on the specifics of
OHe nuclear interaction.
If dark matter can bind to
normal matter, the observations could come from radiative capture of thermalized OHe and could depend on the detector composition and temperature.
In the matter of the underground detector local concentration of OHe is determined by the equilibrium between the infalling cosmic OHe flux and its
diffusion towards the center of Earth. Since the infalling flux experiences annual changes due to Earth's rotation around Sun, this local OHe
concentration possess annual modulations.

 The positive results of the DAMA/NaI and
DAMA/LIBRA
experiments are then explained by the annual modulations of the rate of radiative capture of OHe
by sodium nuclei. Such radiative capture to a low energy OHe-nucleus bound state is possible only for intermediate-mass nuclei:
this explains the negative results of the XENON100 and LUX experiments. The rate of this capture can be calculated by the analogy with
radiative capture of neutron by proton, taking into account the scalar and isoscalar nature of He nucleus, what makes possible only E1 transition
with isospin violation in this process. In the result this rate is
proportional to the temperature (to the square of relative velocity in the absence of local thermal equilibrium): this leads to a suppression of this effect in cryogenic detectors, such as CDMS.

\subsection{OHe solution for positron line excess in the galactic bulge}
The timescale of OHe collisions in the Galaxy exceeds the age of the Universe, what proves that
the OHe gas is collisionless. However the rate of such collisions is nonzero and grows in the regions
of higher OHe density, particularly in the central part of the Galaxy, where these
collisions lead to OHe
excitations. De-excitations of OHe with pair production in E0 transitions can explain the
excess of the positron-annihilation line, observed by INTEGRAL in the galactic bulge \cite{DMRev,DDMRev,CDM,KK2,KMS,CKW,CKW3}.
The calculated rate of collisions and OHe excitation in them strongly depends on OHe density and relative velocity and
the explanation of positron excess was found to be very sensitive to the dark matter density in the central part of Galaxy, where baryonic matter dominates
and theoretical estimations are very uncertain. The latest analysis of dark matter distribution favors more modest values of
dark matter central density, what fixes the explanation of the
excess of the positron-annihilation line by OHe collisions and de-excitation in a very narrow range of the mass of O$^{--}$ near 1.25 TeV.

\subsection{OHe solution for high energy positron excess}
In a two-component dark atom model, based on Walking Technicolor, a sparse WIMP-like component of atom-like
state, made of positive and negative doubly charged techniparticles, is present together with the dominant OHe dark atom and the decays of
doubly positive charged techniparticles to pairs of same-sign leptons can explain the excess of
high-energy cosmic-ray positrons, found in PAMELA and AMS02 experiments \cite{laletin}.
This explanation is possible for the mass of decaying +2 charged particle below 1 TeV and depends on the branching ratios of leptonic channels.
Since even pure lepton decay channels are inevitably accompanied by gamma radiation the important constraint on this model follows from the measurement of cosmic gamma ray background in FERMI/LAT experiment.
The multi-parameter analysis of decaying dark atom constituent model is under way in order to determine the maximal model independent value of the mass
of decaying +2 charge particle, at which this explanation is possible.

\subsection{The LHC probes for OHe solutions for cosmic positron excess}
These astroparticle data can be fitted, avoiding many astrophysical uncertainties of WIMP
models, for a mass of O$^{--}$ $\sim 1$ TeV, which stimulates searches for stable doubly
charged lepton-like particles at the LHC as a test of the composite-dark-matter scenario.
The search for stable multi-charge particles in ATLAS and CMS experiments gives the lower
value for double charged particles around 700 GeV \cite{atlas}. This search will continue in the current Run
of the LHC, giving the hope on the complete experimental test of composite dark matter
explanation for the observed low and high energy positron excess.

\section{Open problems of OHe scenario}
\label{open}
\subsection{The problem of OHe nuclear barrier}

The crucial problem of OHe scenario is the existence of a dipole barrier in OHe nuclear interaction.
The scenario in which such a barrier does not appear was considered in  \cite{CKW4}.
This lead to a significant role of inelastic reactions for OHe, and strongly modified the main features of the OHe scenario.
In the period of Big Bang Nucleosynthesis, when OHe was formed, it captured an additional He nucleus, so that the dominant 
form of dark matter became charged, recombining with electrons in anomalous isotopes of helium and heavier elements.
The resulting over-abundance of anomalous isotopes in terrestrial matter seems to be unavoidable in this case.

This makes the full solution of OHe nuclear physics, started in \cite{CKW2}, vital.
The answer to the possibility of the creation of a dipole Coulomb barrier in OHe interaction with nuclei is crucial.
Without that barrier  one gets no suppression of inelastic reactions, in which O$^{--}$ binds with nuclei. These charged species form atoms (or ions)
 with atomic cross sections, and that strongly suppresses their mobility in terrestrial matter, leading to their storage
 and over-abundance near the
 Earth's surface and oceans.
Hence, the model cannot work if no repulsive interaction appears at some distance between
OHe and the nucleus, and the solution to this question of OHe nuclear physics is
vital for the composite-dark-matter OHe scenario.
\subsection{The problem of the Earth's shadowing}
The terrestrial matter is opaque for OHe, what should inevitably lead to an effect of Earth matter shadowing for
the OHe flux and corresponding  diurnal modulation. This effect needs special study in the confrontation with the constraints, recently
obtained in DAMA/LIBRA experiment  \cite{DAMAshadow}.
\section{Conclusion}\label{conclusion}
The existence of new stable electrically charged particles poses an immediate question on their presence in the surrounding matter in the form of anomalous isotopes, whose possible abundance is severely constrained by the experimental data. The original approach of the sinister model \cite{Glashow} could not overcome the trouble of overproduction of anomalous hydrogen and helium \cite{BKSR1}. However, this approach revealed two important aspects of composite dark matter: possibility of clusters of heavy stable quarks with suppressed QCD interaction and a possibility of stable charged particles hidden in neutral dark atoms.
The development of dark atom scenario during the past decade gave rise to
the OHe composite-dark-matter scenario.

The advantages of this scenario is that it is minimally related
to the parameters of new physics and is dominantly based on the effects of known atomic
and nuclear physics.
However, the full quantum treatment of this problem turns out to be rather
complicated and remains an open (see \cite{CK} for the most recent review).

At the mass of long living stable double charged particles near 1 TeV dark atom scenario can explain the observed excess of low energy positrons in
the galactic bulge and the excess of high energy positrons above 10 GeV, challenging experimental search for such particles at the LHC.

\section*{Acknowledgements}
The work was performed within the framework of the Center
FRPP supported by MEPhI Academic Excellence Project (contract 02.03.21.0005, 27.08.2013). The part on initial cosmological conditions was supported by the Ministry of Education and Science of Russian Federation,
project 3.472.2014/K  and on the forms of dark matter by grant RFBR 14-22-03048.

\end{document}